\documentclass[twocolumn,showpacs,preprintnumbers,amsmath,amssymb,aps]{revtex4}

\usepackage{graphicx}
\usepackage{bm}
\usepackage{epsfig}

\newcommand{\be}{\begin{equation}}
\newcommand{\ee}{\end{equation}}
\begin{document}


\title{New thought experiment to test the generalized second law of
       thermodynamics}

\author{George E.\ A.\ Matsas}
\email{matsas@ift.unesp.br}
\author{Andr\'e R. Rocha da Silva}
\email{dasilva@ift.unesp.br}
\affiliation{Instituto de F\'\i sica Te\'orica,
             Universidade Estadual Paulista,
             Rua Pamplona 145, 01405-900,
             S\~ao Paulo, S\~ao Paulo, Brazil}

\date{\today}

\begin{abstract}
We propose an extension of the original thought experiment proposed by
Geroch, which sparked much of the actual debate and interest on black hole
thermodynamics, and  show that the generalized second law of
thermodynamics is in compliance with it.
\end{abstract}
\pacs{04.70.Dy, 04.62.+v}

\maketitle

In 1970 Geroch~\cite{G} raised the possibility of violating the 
ordinary second law of thermodynamics with help of
classical black holes. The idea was to bring {\em slowly}
from infinity a box with proper energy $E_{\rm b}$ over the event 
horizon and  throw it eventually inside the hole.
The cycle would be closed by lifting back the ideal rope, 
which is assumed to have arbitrarily small mass. Because 
static asymptotic observers would ascribe zero energy to
the box at the event horizon, the hole would remain the 
same after engulfing the box. This would challenge 
the ordinary second law of thermodynamics, since eventually
all entropy associated with the box would be vanished from the
Universe with no entropy increase counterpart. 

As an objection to Geroch's process, Bekenstein argued 
that quantum mechanics would constraint the size and 
energy of the box  accordingly. This would prevent the 
box from reaching the event horizon as a whole and, thus, 
the black hole would necessarily gain mass after engulfing 
the box. Then, Bekenstein~\cite{B} conjectured that black 
holes would have a non-zero entropy 
$S_{\rm bh} = k c^3 A /(4 \hbar G)$ proportional to the 
event horizon area $A$  
and formulated the 
{\em Generalized Second Law} (GSL), namely, that the 
total entropy of a closed system (including that
one associated with black holes) would never decrease. 
Now, because the GSL would be violated when the box entropy 
satisfied $S> 2 \pi k E_{\rm b} R/ (\hbar c)$, where $R$ is the 
proper radius of the smallest sphere which 
circumscribes the box~(see Ref.~\cite{Bo} for a comprehensive 
discussion), Bekenstein conjectured in addition
the existence of a new thermodynamical law, namely, that every system
should have an entropy-to-energy ratio satisfying 
$S/E_{\rm b} \leq 2 \pi k R/(\hbar c)$. 

Notwithstanding, in 1982 Unruh and Wald showed~\cite{UW} that
by taking into account the buoyancy force induced by the Hawking
radiation~\cite{Ha}, as a comprehensive semiclassical gravity
analysis  would demand (notice that $S_{\rm bh}$ depends 
on $G,c$ and $\hbar$), the GSL would {\em not} be 
violated irrespective of the imposition 
of the constraint $S/E_{\rm b} \leq 2 \pi k R/ (\hbar c)$. 
The thermal ambiance outside the hole would prevent 
the box from descending 
beyond the point after which the energy delivered to the black 
hole would be too small to guarantee $\delta S_{\rm bh} \geq S$ 
as demanded by the GSL.

Unruh and Wald's resolution depends crucially on the precise point 
where the box finds its
hydrostatic equilibrium: were it {\em lower}, the GSL would be 
violated. This circumstance led us to analyze an extension of 
the Geroch process in which the box is given some angular 
momentum before it enters the hole. In this case, one can
decompose the force on the box into four distinct components. The
first two ones correspond to the gravitational and buoyancy
forces, which are already present when the box is static outside 
the hole. The remaining ones correspond to the centrifugal 
force and to an extra one, denominated here {\em kinetic gravitational 
force} (see Ref.~\cite{M}), which effectively increases the 
gravitational force on the box. Close enough to the black hole, 
i.e., $r < 3 GM/c^2$, the
kinetic gravitational force  surpasses the centrifugal
one~\cite{AL}, and the equilibrium point is lower
than when the box is at rest. Thus, to rescue the GSL
we must rely on the box's kinetic energy, which is the single 
new ingredient
added to the original Geroch process. Indeed, we show 
here that, the kinetic energy given 
to the box increases enough its total energy to compensate the
reduction of the potential energy caused by the lowering
of the equilibrium point. In this way, the energy given to the hole
is enough to guarantee $\delta S_{\rm bh} \geq S$. 
The precise increase of
the total entropy in this process is displayed. We use natural units 
$c=\hbar=G=k=1$ throughout the rest of the paper. 


Let us describe our static black hole by the line element
\begin{equation}
ds^2 = -\chi^2 dt^2 + \chi^{-2} dr^2
+ r^2 (d\theta^2 + \sin^2 \theta d\phi^2) \;,
\label{ss}
\end{equation}
where $ \chi =\sqrt{1-2M/r} $ is the gravitational redshift factor.
The hole which is assumed to be in thermal equilibrium with Hawking 
radiation can be thought as being enclosed in a large container made of
adiathermal walls~\cite{SH}. 

Our thermodynamical analysis will be carried out 
by Killing observers at rest with the thermal radiation which is 
treated as a perfect fluid (see Refs.~\cite{W}-\cite{FMW} 
and references therein for a recent discussion). 
This is characterized by the stress-energy tensor
\begin{equation}
T^{\mu \nu} =
        (e+ p)u^\mu u^\nu +
         p g^{\mu \nu}  \;,
\label{seT}
\end{equation}	 
where $e=e(r)$ and $p=p(r)$ are the proper energy density 
and pressure, respectively, and 
$u^{\mu}= \chi^{\mu}/\chi$ is the corresponding 
4-velocity with $\chi^\mu = (\partial_t)^\mu$. The associated
proper acceleration 
$
a_{\rm s}= \sqrt { a_{\rm s}^\alpha a^{\rm s}_\alpha }
$
(with $ a^\alpha_{\rm s} = u^\nu \nabla_\nu u^\alpha$)
is 
\begin{equation}
a_{\rm s}= M/\chi r^2 \;.
\label{as}
\end{equation}
From the condition
$
\nabla_\mu T^{\mu \nu} = 0
$, 
we obtain 
$
\nabla^\mu p + (e+p)a^{\mu}_{\rm s}=0
$, 
which leads to 
\begin{equation}
e {d\chi}/{ dl} + { d(\chi  p)}/{dl} =0  \;\;,
\label{hee}
\end{equation}
where
$
l(r) \equiv \int^{r}_{2M}dr'/\sqrt{1-2M/r'}
$.

For the sake of simplicity, we assume that 
our box is rectangular, has proper volume $V$
and is {\em thin}, i.e.
$\delta l \; d\chi / dl \ll \chi$ everywhere in the box,
where $\delta l$ is the box's proper height. 
This condition will be not only physically desirable
as a way to minimize {\em turbulence} and {\em shear} effects 
but also technically convenient as will be seen further.

\begin{figure}[t]
\fbox{\epsfig{file=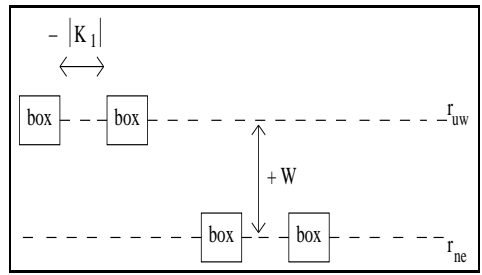, angle=0, width=6cm, height=3.3cm}}
\caption{\footnotesize{A sketch of our thought experiment
is depicted above. Note that 
for $r_{\rm uw} < 3M$, we have $r_{\rm ne}<r_{\rm uw}$.
\label{figure1}}}
\end{figure}

The process which we consider here is as follows. 
Firstly the box is lowered slowly from infinity by some agent
towards the 
black hole up to the point  where $r \equiv r_{\rm{uw}}$, 
in which place it finds its hydrostatic  equilibrium. 
As shown by Unruh and Wald~\cite{UW}, $r_{\rm{uw}}$ is the 
solution of the equation $E_{\rm b} = V e$. In this step some work 
$W_{\rm uw}>0$ is gained by the asymptotic static agent. Now, 
he/she spends some energy to put the box in uniform circular 
motion with angular velocity 
$\omega_0 =d\phi/dt={\rm const}$ (at $\theta = \pi/2$). 
We argue further that this can be done without 
significantly disturbing  the background radiation.
The energy spent by the agent in this part of the 
process  is  denoted by $K_1$, where $K_1<0$ in our convention. 
Clearly, the
hydrostatic equilibrium point changes as the consequence of the
motion (see Fig.~\ref{figure1}). In the process of bringing the box 
to its new equilibrium point the asymptotic agent gains some extra 
work $W>0$, where we assume here that the angular momentum 
$J$ is kept constant. Next, we suppose that the box is 
released and allowed to fall into the black hole, which is supposed to remain 
in equilibrium with its thermal atmosphere~\cite{GW}. 
Any entropy increase in the dropping 
process will be disregarded here because we are interested 
in analyzing the most challenging situation for the 
GSL in the context of rotating boxes, i.e. the one where 
the final total entropy is the least. This is also the reason
why we release the box at the equilibrium point, since this is
where the minimum amount of energy is delivered to the black 
hole. At the end, the angular momentum and energy delivered to 
the hole are $\delta L = J$ and
\begin{equation}
\delta M = E_{\rm b}-W_{\rm{net}}\;,
\label{ed}
\end{equation}
respectively, where
$
W_{\rm{net}}=W_{\rm{uw}}+K_1 + W
$
is the net work gained by the asymptotic agent. 

Because we assume $E_{\rm b}\ll M$, we only consider 
first-order terms
in the expression for the black hole entropy 
increase
\begin{equation}
\delta S_{\rm{bh}}= \frac{\delta M}{T_{\rm{bh}}} 
                  = \frac{E_{\rm b} - W_{\rm uw}}{T_{\rm bh}} 
                   + \frac{|K_1| - W}{T_{\rm bh}}  \;,
\label{Sbh}		  
\end{equation}
where $T_{\rm bh} = 1/(8\pi M)$ (see Refs.~\cite{RG} and \cite{RMW}). 
This was obtained by differentiating 
$S_{\rm bh} = S_{\rm bh} (E,L) = 2\pi E^2 (1+\sqrt{1-L^2/E^4\;})$
around $E=M$ and $L=0$ and using  Eq.~(\ref{ed}).
Now assuming the most challenging case, where 
the box is filled with thermal radiation, it can be shown
that $(E_{\rm b} - W_{\rm uw})/T_{\rm bh} = S_{\rm
b}$~\cite{UW} (see also Refs.~\cite{W} and \cite{BUBW}). 
In our process with the moving box, the change in the generalized 
total entropy is, thus,
\begin{equation}
\delta S_{\rm{g}}=\delta S_{\rm{bh}}-S_{\rm b} 
                 =(|K_1| -W)/{T_{\rm{bh}}}\;.
\label{Sg}		 
\end{equation}
As a result, the GSL will be satisfied depending whether or not 
$|K_1| -W\geq 0$.  
In order to decide on it, we have to analyze more carefully the subprocess, 
where the box gains the kinetic energy $K_1$ and moves towards its new 
equilibrium point ($r=r_{\rm ne}$) along which the asymptotic agent gains 
the work $W$. (Naturally, if the box is not put in motion, $|K_1| = W = 0$ and
we recover Unruh and Wald's result.)   
At this point, we would like to make two remarks about the process of 
setting the box in motion. First, we do not 
want that the moving box disturbs much the thermal atmosphere 
because the associated entropy increase would be difficult to compute. 
This should be partly achieved by using {\em thin} boxes or
by considering a set of boxes rather than a single one. They would be 
lowered from infinity to $r_{\rm uw}$ and fitted one with the other 
forming a closed ring around the black hole. This would eliminate front 
and rear particle shocks with the box walls, 
which would disrupt the energy distribution (and entropy) of the 
thermal bath. Particle shocks with the up 
and down walls (which would still exist) are not source of concernings, 
since they do not lead to energy or momentum transfer.  
In this paper, the assumption of a single thin box will suffice. 
After all, the existence of other 
sources of entropy increase would only help to render the GSL valid. 
Now, the use of thin boxes is also useful to solve our second concerning.  
In order to keep the box uncorrupted during the initial acceleration 
interval, we impose that the 4-velocity
$v^\mu$ of the box's points satisfy the {\em no expansion
condition}: 
$ 
\Theta\equiv \nabla_\mu v^\mu = 0
$. 
This can be realized by choosing
$
v^\mu(x^\alpha) =\left[\chi^\mu + \omega(x^\alpha) \phi^\mu \right]/
                |\chi^\mu + \omega(x^\alpha) \phi^\mu)| 
$
with 
$
\phi^\mu = (\partial_{\phi})^{\mu}
$
and 
$
\omega(x^\alpha)=\chi^{2}t/r^{2}\phi \leq \omega_{0}
$
for
$
0 \leq t/\phi \leq \omega_0 r^2 / \chi^2
$,
where
$
0 < \omega_{0} < \chi/r
$. 
This  is  necessary but not sufficient 
to guarantee the validity of the rigid body condition
$
\sigma_{\mu \nu} + (\Theta/3) h_{\mu \nu} = 0
$,
i.e., that the proper distance among the box's points
are kept the same, where 
$ 
h_{\mu \nu} \equiv g_{\mu \nu} + v_{ \mu } v_{ \nu } 
$ 
and
$
\sigma_{\mu \nu} \equiv
h^{\;\alpha}_{\mu}h^{\;\beta}_{\nu}\nabla_{(\alpha}v_{\beta)}- 
(\Theta/3) h_{\mu \nu}
$
is the shear tensor. Happily, however, the use of our
thin box assumption leads to an approximate verification 
of the rigid body condition (see Ref.~\cite{M} 
for a comprehensive discussion). (The thinner the box, 
the more the rigid body equation is satisfied.) Finally, we 
also stress that as the box reaches its uniform circular 
motion the rigid body
condition is fully verified and no distortion appears at all. 

In order to compute Eq.~(\ref{Sg}), we begin recalling that 
in the uniform motion regime, 
$
t/\phi >\omega_{0}r^2/\chi^2
$,
the box's points 
have 4-velocity 
$
v^\mu = \eta^{\mu}/\eta
$,
where
$
\eta^\mu = \chi^{\mu}+\omega_{0}\phi^{\mu}
$
with
$
\eta = \sqrt{\chi^{2}-r^{2}\omega^{2}_{0}}
$
and proper acceleration 
$
a_{\rm m}= \sqrt{ a_{\rm m}^\alpha a^{\rm m}_\alpha }
$
(with $a_{\rm m}^\alpha = v^\nu \nabla_\nu v^\alpha$),
which can be rewritten as 
\begin{equation}
a_{\rm m} = \eta^{-1}d\eta/dl\;.
\label{am}
\end{equation}
The box's angular momentum and kinetic energy 
at $ r= r_{\rm uw}$ as defined asymptotically are 
\begin{eqnarray}
J&\equiv& E_{\rm b} v^{\mu}\phi_{\mu}|_{r=r_{\rm uw}} 
\nonumber \\
&=& E_{\rm b} \omega_0 r^2/\eta |_{r=r_{\rm uw}} 
\label{angular} 
\end{eqnarray}
and
\begin{eqnarray}
K_{1} &\equiv & E_{\rm b}[\chi^{\mu}\left(u_{\mu}- v_{\mu} \right)]_{r=r_{\rm uw}} 
\nonumber \\    
& = & - E_{\rm b}|\chi \left(1-\chi/\eta\right)|_{r=r_{\rm uw}} \;.
\label{K1}
\end{eqnarray}  
The local force on the box is
\begin{eqnarray}
F_{\rm{loc}} 
& \equiv & E_{\rm b} a_{\rm m} 
\label{aJoriginal} \\
&=&
\frac{ME_{\rm b}}{\chi r^{2}}-
\frac{\chi J^{2}}{E_{\rm b}r^{3}}+
\frac{M J^{2}}{\chi E_{\rm b}r^{4}}\;.
\label{aJ} 
\end{eqnarray}
The first two terms in the right-hand side should be identified 
with the gravitational force on the box when it lies at rest and
with the centrifugal force, respectively. The last term (which involves
$M$ as well as $J$) is what we have called kinetic gravitational force. 
By using Eq.~(\ref{angular}), one can verify 
that for $r < 3M$ the kinetic gravitational force 
is larger than (the absolute value of) 
the centrifugal force. As a result, for $r<3M$ 
the {\em new} equilibrium point for the {\em moving} box will be closer 
to the black hole, i.e., $r_{\rm ne} < r_{\rm uw}$ (see Fig.~\ref{figure1}).  

In order to obtain $r_{\rm ne} $, we must calculate the 
buoyancy force on the moving box. The proper hydrostatic 
pressures on the top ($r=r_\top$) and at the bottom ($r=r_\bot$) 
of the box are
$
P_{\top/\bot} \equiv T_{\mu \nu} n^{\mu}_{\top/\bot} n^{\nu}_{\top/\bot}
= p(r_{ \top/\bot})  
$,
where 
$
n^{\mu}_{ \top/\bot }=\chi(r_{\top/\bot })(\partial_{r})^{\mu}
$
are unit vectors orthogonal to the box's 4-velocity.
Consequently, the hydrostatic scalar forces on the top and at the bottom 
of the box are
$
F_{\top}= - Ap_{\top}
$ 
and
$
F_{\bot}= Ap_{\bot}
$,
respectively, where $A$ is the corresponding proper area.
In order to obtain the buoyancy force, we  transmit
both $F_{\top}$ and $F_{\bot}$ to the point ${\cal{O}}$,
where the local force~(\ref{aJ}) is calculated. Let us assume that the 
forces are transmitted through {\em ideal} cables and rods 
characterized by the stress-energy tensor
$
{\cal{T}}^{\mu \nu} = P_{\rm c/r} h^{\mu \nu}
$ 
satisfying 
$
\nabla_{\mu}{\cal{T}}^{\mu \nu}=0
$,
where $P_{\rm c/r}$ stands for pressure. 
Thus, from $F_{\top/\bot}$ we obtain the transmitted forces 
$F^{\cal{O}}_{\top/\bot}$ at ${\cal{O}}$ as
$
F_{\top/\bot}^{\cal O} =
[\eta({r}_{\top/\bot})/\eta({r}_{\cal O})]F_{\top/\bot}
$.
The buoyancy force is, then, written as
\begin{equation}
F^{\cal O}_{\rm buo}
  =  F_{\top}^{\cal O} + F_{\bot}^{\cal O}
  =  \left. \frac{V}{\eta}
       \frac{d(\eta p)}{dl}
        \right|_{{r} = {r}_{\cal O}}  \;,
\label{Fbuo}
\end{equation}
where we have used our thin box assumption, 
namely, that
$
\delta l \, d(\eta p)/dl \ll \eta p
$ 
everywhere in the box so that we can neglect higher derivative
terms in Eq.~(\ref{Fbuo}).
\begin{figure}[t]
\fbox{\epsfig{file=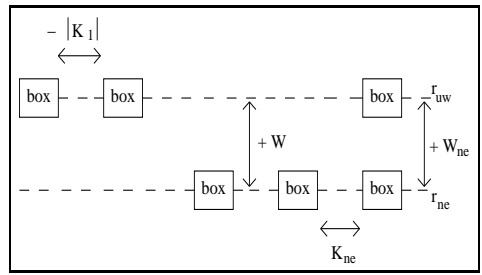,angle=0, width=6cm, height=3.3cm}}
\caption{\footnotesize{A sketch of our auxiliary 
closed cycle is depicted above.}
\label{figure2}} 
\end{figure}

Now, by adding up Eqs.~(\ref{aJoriginal})  
and (\ref{Fbuo}) we obtain the total local force on the box as 
\begin{equation}
F^{\cal O}_{\rm tot} =
V\left[ \frac{\rho_{\rm{b}}}{\eta} \frac{d\eta}{dl}
+ \frac{1}{\eta} \frac{d(\eta p )}{dl}\right]_{{r} = {r}_{\cal O}}\;,
\label{Flt}
\end{equation}
where $\rho_{\rm{b}}=E_{\rm b}/V$. Now, we must note that the 
corresponding total local 4-force points along $(\partial_r)^\mu$, and so 
it also 
lies in the spacelike section of the static observers. 
As a result, the static observers ascribe the same 
force $F^{\cal O}_{\rm tot}$ acting on the box. Hence, 
the force which the asymptotic agent must apply to sustain the box 
is
$
F^{\infty}_{\rm{tot}}=\chi(r_{\cal O}) F^{{\cal{O}}}_{\rm{tot}}
$, 
which can be recast in the form [see Eq.~(\ref{hee})]
\begin{equation}
F^{\infty}_{\rm{tot}}=
V 
\left[
\frac{M ( \rho_{\rm b} -e )}{r^2} +
\frac{J^2 ( \rho_{\rm b} + p )}{E_{\rm b}^2 r^3 } 
\left(\frac{3M}{r}-1 \right) 
\right]_{{r} = {r}_{\cal O}} \;.
\label{Fit}
\end{equation}
Clearly in the limit where 
$
J \rightarrow 0
$,
this expression is equal to Unruh and Wald's result 
\begin{equation}
F^{\infty}_{\rm{uw}}= V (\rho_{\rm b }- e  )\chi a_{\rm{s}}
\label{Fuw}
\end{equation}
(see Refs.~\cite{UW} and \cite{PW}). 
Note that if $ r_{\rm{uw}} =r_{\cal O} <3M$, then $F^{\infty}_{\rm{tot}}>0$ 
(where we recall that $F^{\infty}_{\rm{uw}}=0$) and the box is 
pulled downwards. The new equilibrium point at $r= r_{\rm ne}$ 
is obtained as the solution of $F^{\infty}_{\rm tot}(r_{\rm ne})=0$, i.e.
\begin{equation}
[M(\rho_{\rm b}-e) \eta^2 (r_{\rm uw})\, r^2 +
\omega^{2}_{0} (\rho_{\rm b} + p)r^{4}_{\rm uw}(3M-r)]_{r=r_{\rm ne}}=0
\;,
\label{rne}
\end{equation}
where the radial dependence of $p=p(e)=p[e(r)]$ is required.

The work $W$ gained by the asymptotic agent as the box is lowered from
$r=r_{\rm uw}$ to $r=r_{\rm ne}$ is 
\begin{equation}
W = -\int^{r_{\rm{ne}}}_{r_{\rm{uw}}} F^{\infty}_{\rm{tot}}dr/\chi \;,
\label{Work}
\end{equation}
where $ F^{\infty}_{\rm{tot}}$ is given in Eq.~(\ref{Fit}).

Before using Eqs.~(\ref{K1}) and~(\ref{Work}) in Eq.~(\ref{Sg}) to calculate  
explicitly $\delta S_{\rm g}$, let us use first a shortcut to show  
that $\delta S_{\rm g}>0$. For this purpose, let us add two extra steps in
our original cycle 
as follows. Rather than throwing the box to the black hole at $r=r_{\rm ne}$, 
we (i) stop the box and (ii) bring it back to $r=r_{\rm uw}$ 
(see Fig.~\ref{figure2}). In the process of stopping it, the asymptotic 
agent gains an energy
\begin{equation}
K_{\rm{ne}}= E_{\rm b} |\chi (1-\chi/\eta) |_{r=r_{\rm{ne}}} 
\label{Kms}
\end{equation}  
and in the process of pulling it back from $r_{\rm ne}$ to 
$r_{\rm uw}$, he/she  also gains an extra energy 
\begin{equation}      
W_{\rm ne}=-\int^{r_{\rm{uw}}}_{r_{\rm_{ne}}}F^{\infty}_{\rm{uw}}dr/\chi\;,
\label{Wpullb}
\end{equation}
where 
$F^{\infty}_{\rm uw}$ is given in Eq.~(\ref{Fuw}).  
By assuming that the closed cycle which brings the box from 
$r_{\rm uw}$ to $r_{\rm ne}$ and back to  $r_{\rm uw}$ is conservative, 
we must have
$K_1 + W + K_{\rm{ne}}+W_{\rm ne} =0 $,
i.e.
\begin{equation}
|K_1| -W =K_{\rm{ne}}+W_{\rm ne}\;.
\label{identity}
\end{equation} 
Then from Eq.~(\ref{Sg}), we obtain
\begin{equation}
\delta S_{\rm{g}} = (K_{\rm{ne}}+W_{\rm ne})/T_{\rm{bh}}>0 \;.
\label{final}
\end{equation}
This guarantees that the box's energy increase of kinetic origin $|K_1|$ 
is enough to compensate the energy decrease of gravitational origin $W$  
[see Eq.~(\ref{Sg})], saving the GSL.
\begin{figure}[t]
\epsfig{file=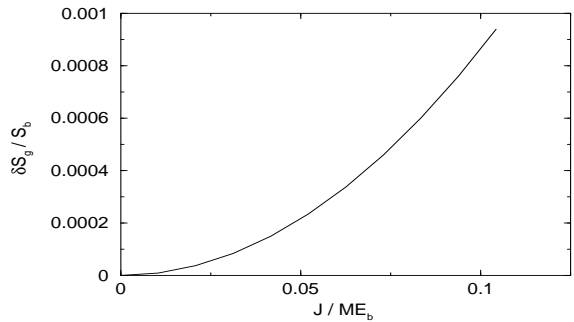, angle=0, width=7.5cm, height=4.3cm}
\caption{\footnotesize{Here we plot the generalized total 
entropy increase $\delta S_{\rm g}$ as a function of the 
box angular momentum $J$.}
\label{figure3}}
\end{figure}

Now, we proceed to calculate explicitly $\delta S_{\rm{g}}$.  
For this purpose, we assume (for simplicity ) that 
the Hawking radiation and the box only contain
a single free massless bosonic field, say, photons, 
in which case  $p=e(r)/3$ with $e= (\pi^{2}/15) T^{4}$ 
and $T=T_{\rm bh}/\chi$ is the Tolman's relation~\cite{To}.
In this case 
\begin{equation}
r_{\rm uw}= \frac{2M}{1- \sqrt{(\pi^{2}T_{\rm bh}^{4}) /(15 \rho_b)}} \;,
\label{ruw}
\end{equation}
where we impose 
$\rho_{\rm b} > 9 \pi^2 T_{\rm bh}^4/15 $ 
to guarantee that 
$2M < r_{\rm uw} < 3M$ [and we recall that $r_{\rm ne}$ is given 
in Eq.~(\ref{rne})]. 
Finally, we are in position to evaluate numerically 
$\delta S_{\rm{g}}$. As a check, we use independently 
Eqs.~(\ref{Sg}) and~(\ref{final}). The results are plotted
in Fig.~\ref{figure3}. 

The existence of Hawking radiation has allowed us to ascribe
temperature to black holes. This in addition with the laws of
black hole mechanics led us to associate entropy to these
objects. However, in order to treat black holes as legitimate
thermodynamical systems it is necessary to conjecture the
GSL. Since it is not possible to develop direct
tests for the GSL, the best we can do is to verify its validity 
through thought experiments devised in contexts, where 
our well known theories can be safely used. In these vein, we 
have offered here a new thought experiment and shown that the 
GSL complies with it.

\begin{acknowledgments}

G.M. acknowledges partial support from Conselho Nacional de
Desenvolvimento Cien\-t\'\i fico e Tecnol\'ogico and Funda\c c\~ao
de Amparo \`a Pesquisa do Estado de S\~ao Paulo and
A.S. acknowledges full support from Funda\c c\~ao
de Amparo \`a Pesquisa do Estado de S\~ao Paulo.

\end{acknowledgments}

\end{document}